\documentclass[aps,prb,letterpaper,amsmath,amssymb,superscriptaddress,reprint,
]{revtex4-1}
\usepackage{graphicx}

\begin{document}
\title{Large exchange bias in polycrystalline MnN/CoFe bilayers at room temperature}
\author{Markus Meinert}
\email{meinert@physik.uni-bielefeld.de}
\author{Bj\"orn B\"uker}
\author{Dominik Graulich}
\author{Mareike Dunz}
\affiliation{Center for Spinelectronic Materials and Devices, Bielefeld University, D-33501 Bielefeld, Germany}
\date{\today}

\begin{abstract}
We report on the new polycrystalline exchange bias system MnN/CoFe, which shows exchange bias of up to 1800\,Oe at room temperature with a coercive field around 600\,Oe. The room temperature values of the interfacial exchange energy and the effective uniaxial anisotropy are estimated to be $J_\mathrm{eff} = 0.41\,\mathrm{mJ}/\mathrm{m}^2$ and $K_\mathrm{eff} = 37\,\mathrm{kJ}\,/\,\mathrm{m}^3$. The thermal stability was found to be tunable by controlling the nitrogen content of the MnN. The maximum blocking temperature exceeds $325^\circ$C, however the median blocking temperature in the limit of thick MnN is $160^\circ$C. Good oxidation stability through self-passivation was observed, enabling the use of MnN in lithographically defined microstructures. As a proof-of-principle we demonstrate a simple GMR stack exchange biased with MnN, which shows clear separation between parallel and antiparallel magnetic states. These properties come along with a surprisingly simple manufacturing process for the MnN films.
\end{abstract}

\maketitle

\section{Introduction}
Spinelectronics is becoming an increasingly important technology for the realization of nonvolatile, fast, low-power computer memory and is already well-established in hard disk drive read heads and magnetic sensors.\cite{Gregg2002, Chappert2007} The key component in spinelectronic devices, a magnetoresistive element using either the giant magnetoresistance (GMR) or tunnel magnetoresistance (TMR), is comprised of two magnetic films: a free sense-layer and a fixed reference layer. The ferromagnetic free layer is free to follow external magnetic fields or can be switched by a current via the spin transfer torque. The reference layer has to be stable against external fields to allow for different magnetic alignments of the two layers, which give rise to the magnetoresistance. Usually the reference layer is made by pinning a thin ferromagnetic film (FM) to an antiferromagnetic film (AFM). The exchange bias\cite{Meiklejohn1957,Nogues1999,Berkowitz1999,Stamps2000,Kiwi2001,Ohldag2003,OGrady2010} (EB) associated with this pinning shifts the magnetic hysteresis of the reference layer to fields that are not encountered during normal device operation. The antiferromagnet has to meet a number of criteria to be suitable for integration into such stacks: (1) Temperature stability: the temperature at which the exchange bias of the AFM/FM stack vanishes (the so-called blocking temperature) has to be significantly larger than the device operation temperature. (2) Exchange bias and coercive fields: the exchange bias field has to be significantly larger than the coercive field to allow for a clear separation of the parallel and antiparallel magnetic states of the GMR or TMR stack. (3) Ease of manufacturing: processing with industry-standard magnetron sputtering onto Si wafers is desired, hence polycrystalline systems have to meet the above criteria. High annealing temperatures above 300$^\circ$C and deposition at elevated temperature should also be avoided for integration with semiconductor technology. For lithographic processing, the material needs to be sufficiently stable against oxidation. (4) Environmental safety and price: for large-scale application the price of the material will play a crucial role. Also, use of unsafe or otherwise critical materials should be avoided. (5) For current-in-plane GMR devices, the AFM has to be highly resistive, in order not to short-cut the active GMR stack.

The criteria 1 to 3 are mostly met by the commonly used antiferromagnets PtMn and IrMn,\cite{Nozieres2000,Ali2003,Rickart2005} although PtMn requires high temperature annealing to form the antiferromagnetic phase. However, criterion 4 is obviously violated: Pt and Ir are rare, expensive, and mining for them causes considerable environmental pollution.\cite{Glaister2010} Thus, alternatives to these noble-metal based systems are needed. Other antiferromagnets, such as FeMn and NiMn have poor corrosion resistance or the ratio of exchange bias and coercive fields is not ideal for many applications.\cite{Nozieres2000}

In the present article we report on a new exchange bias system that is based on the antiferromagnet MnN. It crystallizes in the $\theta$-phase of the Mn-N phase diagram,\cite{Gokcen1990} which has the tetragonal face-centered variant of the NaCl structure with $a = 4.256$\,\AA{} and $c = 4.189$\,\AA{} at room temperature (RT), where the precise numbers depend on the N content in the material. With increasing nitrogen content, larger lattice constants are observed.\cite{Otsuka1977,Suzuki2000,Leineweber2000}

The N\'{e}el temperature of MnN is about 660\,K;\cite{Leineweber2000} the magnetic transition is accompanied by a tetragonal-to-cubic structural transformation caused by magnetostriction.\cite{Otsuka1977} The magnetic order of the material was investigated with neutron powder diffraction\cite{Leineweber2000, Suzuki2001} and by first principles calculations.\cite{Lambrecht2003} It was found to be collinear of AFM-I type, i.e. with the magnetic moments coupled parallel within the $c$-planes and alternating along the $c$-direction. The spin orientation is under debate; Leineweber \textit{et al.} found it to be along the $c$-direction just below the N\'{e}el temperature, while it would tilt slightly away from the $c$-axis at lower temperature.\cite{Leineweber2000} Instead, Suzuki \textit{et al.} found the spin direction to be in the $c$-planes.\cite{Suzuki2001} An important difference between these experiments lies in the preparation procedures, that led to a slightly nitrogen-poor $\theta$-MnN in the first case, while the $\theta$-MnN was saturated in the second case, which is also reflected by the larger lattice constants in the latter case. We therefore propose that the data measured by Suzuki \textit{et al.} reflect the intrinsic properties of the stoichiometric $\theta$-MnN phase. In both cases, the Mn magnetic moments were found to be $3.3\,\mu_\mathrm{B}$ at room temperature.

\section{Experimental procedure}

To investigate the suitability of MnN as an antiferromagnet for exchange bias applications, we prepared film stacks of Ta 10\,nm / MnN $t_\mathrm{MnN}$ / Co$_{70}$Fe$_{30}$ $t_\mathrm{CoFe}$ / Ta 0.5\,nm / Ta$_2$O$_5$ 2\,nm on thermally oxidized Si wafers by (reactive) dc magnetron sputtering at room temperature. The sputtering system used here was a 2-inch co-sputtering system with a target-to-substrate distance of 90\,mm and an unbalanced magnetron configuration. The base pressure of the system was around $5 \cdot 10^{-9}$\,mbar prior to the deposition runs. The MnN films were sputtered from an elemental Mn target in a mixed Ar and N$_2$ atmosphere with various working pressures and partial pressure ratios. Optimization of the deposition parameters with respect to the exchange bias yielded a 50:50 N$_2$:Ar mixture at $p=2.3\cdot10 ^{-3}$\,mbar as the best deposition condition. This condition was used unless stated otherwise. The typical deposition rate of MnN was $0.1\,$nm/s at a source power of 50\,W. Subsequent post annealing and field cooling in a magnetic field of $H_\mathrm{FC} = 6.5\,\mathrm{kOe}$ was done in a vacuum furnace.
 
 Magnetic characterization of the stacks was performed using the longitudinal magneto-optical Kerr effect (MOKE) at room temperature. The magnetic field was applied in the film plane in all measurements. Structural and film thickness analysis was done with a Philips X'Pert Pro MPD, which is equipped with a Cu source and Bragg-Brentano optics. For in-plane measurements an open Euler cradle and point focus optics were used.

\section{Results}

\subsection{Crystal Structure}
\begin{figure}[h!]
\includegraphics[width=8.5cm]{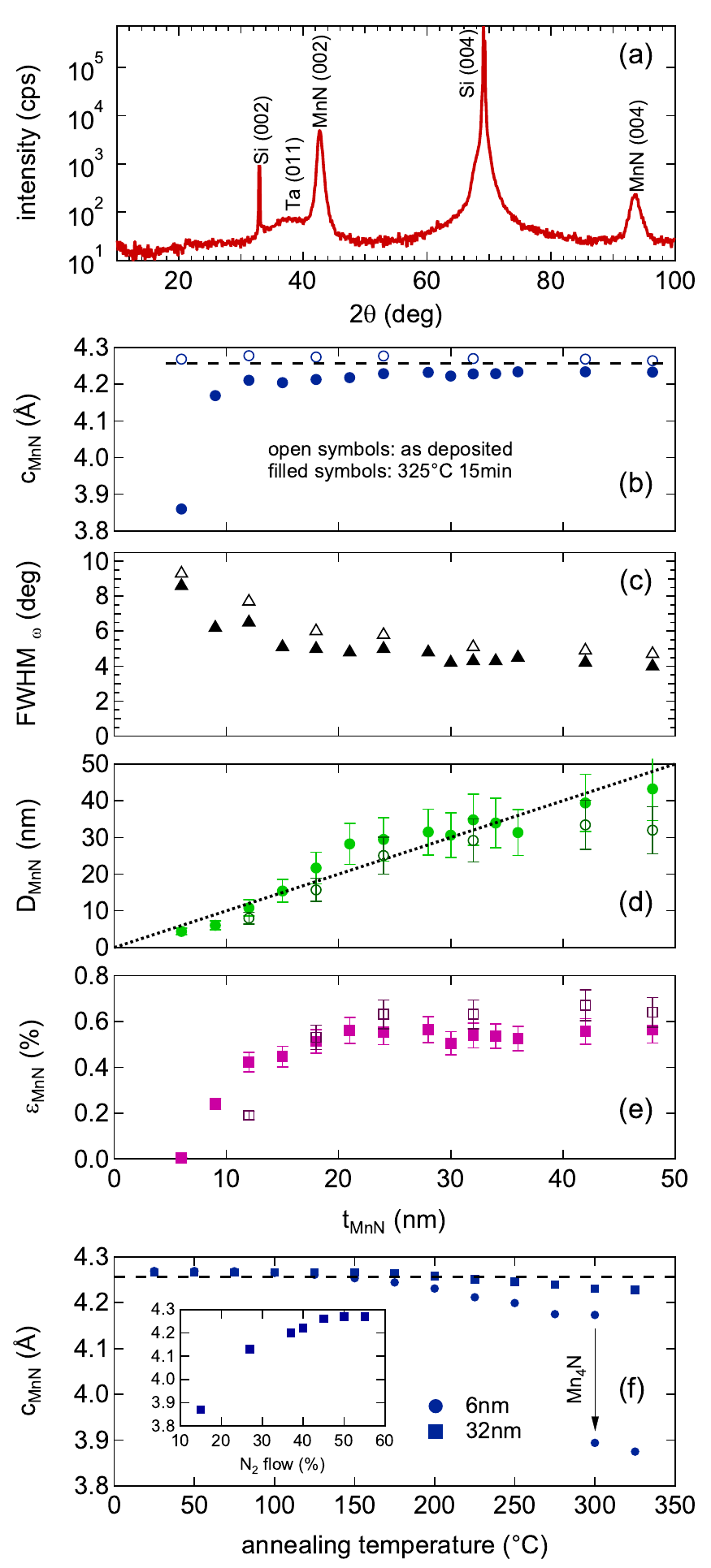}
\caption{(a): X-ray diffraction spectrum of a Ta\,10\,/\,MnN\,32 film annealed at 300$^\circ$C for 15\,min. (b): Lattice constants, (c): full width at half-maximum (FWHM) of the rocking curves of the (002) peaks, (d): perpendicular grain sizes, and (e): microstrain of as-deposited and annealed films as functions of the MnN thickness. (f): Lattice constants as functions of annealing temperature for two MnN film thicknesses. The inset shows the lattice constant as a function of the N$_2$ flow in the sputtering gas. The dashed line in (b) and (f) represents the lattice constant $a$ from Ref. \onlinecite{Suzuki2000}}
\label{fig:xrd_all}
\end{figure}

To verify the growth of the $\theta$-phase of MnN we have performed extensive x-ray diffraction analysis of MnN films with varying film thicknesses and annealing conditions. In Figure \ref{fig:xrd_all}\,(a) a typical x-ray diffraction spectrum of the Ta\,/\,MnN\,/\,CoFe stacks is shown. All peaks are identified as belonging to the substrate, to the Ta seed layer, or to the MnN film. The Ta layer has (011) orientation with very small grains with grain size $D \approx 1\,\mathrm{nm}$. The lattice constant of MnN before annealing is larger than $4.26$\,\AA{}, which is larger than the bulk values reported in the literature. No other phases are observed, and the lattice constant is too large to identify the structure as the cubic $\varepsilon$-Mn$_4$N or the tetragonal $\eta$-Mn$_3$N$_2$ phase.\cite{Yang02} The mass density obtained from x-ray reflectivity is $\rho = (6.1 \pm 0.05)\,\mathrm{g}\,/\,\mathrm{cm}^3$ (without annealing), which is slightly larger than the expected mass density of $6.03\,\mathrm{g}\,/\,\mathrm{cm}^3$ assuming a stoichiometric $\theta$-MnN with the lattice constants given by Suzuki \textit{et al.}\cite{Suzuki2000} This indicates that the in-plane lattice constants are contracted with respect to the literature values. By in-plane diffraction measurements it was found that the lattice constant in the film plane is $a \approx 4.10$\,\AA{}. Thus, the out-of-plane lattice constant is \textit{larger} than the in-plane constant, at odds with the usual $\theta$-phase with $c/a \approx 0.984$. Here, we find $c/a \approx 1.04$. All peaks observed with in-plane diffraction are compatible with an fct phase with $c/a \approx 1.04$. In the following, we shall call this phase $\theta'$-MnN, as it is structurally equivalent to the $\theta$-phase, however it shows a strong tetragonal distortion along the film growth direction. The observed mass density matches the expected mass density of stoichiometric $\theta'$-MnN perfectly.

We present results from a film thickness series in Figure \ref{fig:xrd_all}\,(b)-(e), where films in their as-deposited state and films after annealing at 325$^\circ$C are compared. The perpendicular crystallite size and microstrain contributions to the peak widths were extracted from the (002) and (004) reflections with the Williamson-Hall  equation:\cite{Williamson1953}
\begin{equation}
B_\mathrm{ss}(\theta) \cos\theta = \frac{k \lambda}{D} + 4\varepsilon \sin\theta,
\end{equation}
where $B_\mathrm{ss}$ denotes the size-strain contribution to the peak width after removing the instrumental broadening, $k$ is the Scherrer constant which we take as $k = 0.9$, $D$ is the crystallite size along the film growth direction, and $\varepsilon$ describes the microstrain in terms of a relative lattice constant variation. As the microstrain is fairly large in our films, a straightforward evaluation of the peak widths with Scherrer's formula leads to an erroneous saturation of the perpendicular grain size at 12\,nm. The microstrain can be evaluated with small error bar with this method, however the crystallite size is very sensitive to small errors in the peak width determination, which results in large relative errors. The integral breadths were used to determine the peak widths $B_\mathrm{ss}$ and the instrumental broadening was determined with a strain-free Si pressed powder pellet.

We observe that the lattice constant is independent of the MnN film thickness in the as-deposited state, but after annealing the lattice constant is much smaller for thin films, whereas it shrinks only slightly for thicker films. The rocking curve width is reduced with increasing film thickness and saturates at about $4.8^\circ$ for $t_\mathrm{MnN} > 30\,\mathrm{nm}$. For all film thicknesses the rocking curve width is reduced by about $0.8^\circ$ through annealing at $325^\circ$C for 15\,min, which indicates lateral grain growth and grain boundary crystallization. The Williamson-Hall analysis shows that the perpendicular grain size is directly proportional to the film thickness up to  $t_\mathrm{MnN} \approx 30\,\mathrm{nm}$. Annealing induces further grain growth or defect healing and the perpendicular grain size matches the film thickness up to  $t_\mathrm{MnN} = 48\,\mathrm{nm}$. For thick films, the annealing reduces the microstrain of the films, whereas it is increased for thin films up to  $t_\mathrm{MnN} \approx 15\,\mathrm{nm}$. In both cases it is still rather large, which indicates a large number of defects in the films. In conclusion, we obtained $\theta'$-MnN films with a pronounced $(001)$-fiber texture.


To further investigate the large change in the lattice constant of thin MnN films upon annealing, we made an annealing series with two MnN film thicknesses, the results of which are shown in Figure \ref{fig:xrd_all}\,(f). For a 32\,nm thick film, the lattice constant shrinks slightly with increasing annealing temperature, where the shrinking sets in at about $200^\circ$C. The 6\,nm thick film shows instead a phase transition to a structure with much smaller lattice constant at $300^\circ$C, which we assume to be the $\varepsilon$-Mn$_4$N phase. We suppose that this phase transition arises from N diffusion into the Ta seed layer, which would form a TaN$_x$ layer at the interface and passivate at some point. Thicker MnN films do not show this phase transition because their N reservoir is much larger, so that not enough N can diffuse out of the film to form the $\varepsilon$-Mn$_4$N.

The inset in Figure \ref{fig:xrd_all}\,(f) shows that the lattice constant shrinks with reduced nitrogen flow in the sputtering gas. This arises from the stuffing of the fct Mn lattice with N atoms: with decreasing nitrogen content the lattice is not sufficiently filled and the lattice constant will shrink. At very low N$_2$ flow we find the $\varepsilon$-Mn$_4$N phase again. In the following we will focus on N$_2$ flows larger than 40\%, where the lattice constant indicates a $\theta'$-MnN phase with a small amount of vacancies.

\subsection{Exchange Bias}

\subsubsection{Film Thickness Dependence}

\begin{figure}[t]
\includegraphics[width=8.6cm]{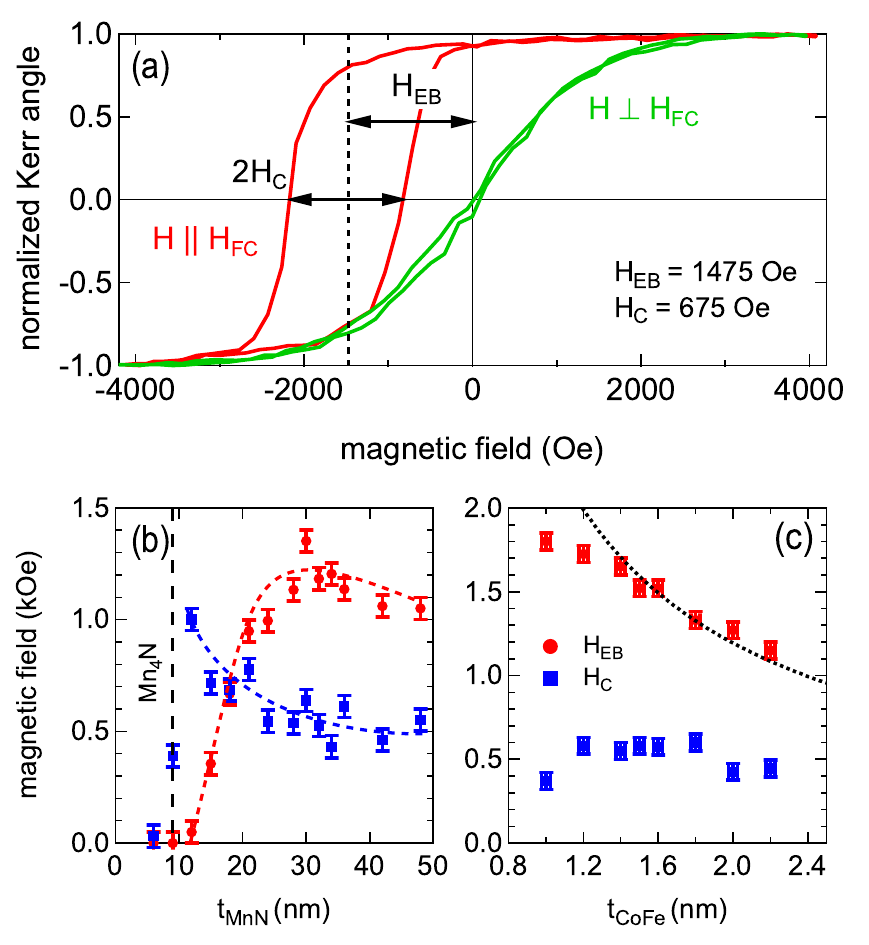}
\caption{(a): MOKE loops parallel and perpendicular (in the sample plane) to the field cooling direction for a sample with $t_\mathrm{MnN} = 32\,\mathrm{nm}$ and $t_\mathrm{CoFe} = 1.6\,\mathrm{nm}$. (b): The dependence of $H_\mathrm{EB}$ and $H_\mathrm{C}$ on the MnN thickness, $t_\mathrm{CoFe} = 1.8\,\mathrm{nm}$. (c): The dependence of  $H_\mathrm{EB}$ and $H_\mathrm{C}$ on the CoFe thickness, $t_\mathrm{MnN} = 30\,\mathrm{nm}$. The samples were annealed at 325$^\circ$C for 15\,min. Dotted lines are guides to the eye.}
\label{fig:moke_loops_thicknesses}
\end{figure}

In Figure \ref{fig:moke_loops_thicknesses}\,(a) we show a magnetic loop of an optimized MnN/CoFe bilayer with $t_\mathrm{MnN} = 32$\,nm and $t_\mathrm{CoFe} = 1.6$\,nm. Our definitions of the exchange bias field $H_\mathrm{EB}$ and the coercive field $H_\mathrm{C}$ are marked with arrows. The loop shows high exchange bias and a reasonably low coercive field with a squareness of $S = M(H_\mathrm{EB})/  M_\mathrm{sat} \approx 0.75$. The ferromagnet is almost saturated at zero external field. In the following we discuss the variation of various stack parameters to identify the optimum deposition and post-annealing conditions for the MnN/CoFe stacks to maximize the exchange bias field and simultaneously minimize the coercive field of the CoFe.

Figures \ref{fig:moke_loops_thicknesses}\,(b) and  \ref{fig:moke_loops_thicknesses}\,(c) display the variation of the exchange bias field $H_\mathrm{EB}$ and the coercive field $H_\mathrm{C}$ with the film thicknesses $t_\mathrm{MnN}$ and $t_\mathrm{CoFe}$. For the MnN thickness variation a maximum of $H_\mathrm{EB}$ was observed at $t_\mathrm{MnN} = 30$\,nm. For $t_\mathrm{MnN} = 6$\,nm and $9$\,nm no exchange bias was found. Between $t_\mathrm{MnN} = 6$\,nm and $t_\mathrm{MnN} = 12$\,nm, the coercive field rises sharply which can be traced back to the phase transition to Mn$_4$N at low MnN film thickness discussed earlier. At $t_\mathrm{MnN} = 12$\,nm the x-ray diffraction indicates that the MnN is preserved upon annealing, so we assume the observed MnN film thickness dependence of the coercive and exchange bias fields to reflect the intrinsic properties of the MnN/CoFe bilayer at and above $t_\mathrm{MnN} = 12$\,nm. Notably, the maximum of the exchange bias is found for significantly larger AFM thickness as compared to IrMn (7\,nm) or PtMn (20\,nm).\cite{Ali2003,Rickart2005}

The CoFe thickness variation, shown in Fig. \ref{fig:moke_loops_thicknesses}\,(c), has the usual $t_\mathrm{CoFe}^{-1}$ dependence for CoFe thickness above $1.4$\,nm. At lower thickness, the exchange bias slightly deviates from this dependence, but still increases with decreasing CoFe film thickness. The hyperbolic part allows to determine the effective interfacial exchange energy within the effective Meiklejohn-Bean model\cite{RaduZabel} $J_\mathrm{eff} = t_\mathrm{CoFe} M_\mathrm{CoFe}  \mu_0  H_\mathrm{EB} = 0.41\,\mathrm{erg}/\mathrm{cm}^2 = 0.41\,\mathrm{mJ}\,/\,\mathrm{m}^2$ using the saturation magnetization of $ M_\mathrm{CoFe} \approx 1700$\,emu/cm$^3$ for our Co$_{70}$Fe$_{30}$ composition.\cite{Reck69} This exchange energy is larger than that of typical IrMn/CoFe stacks and is comparable to that of NiMn/NiFe or PtMn/CoFe.\cite{Nozieres2000,Rickart2005} Chemically well ordered Mn$_3$Ir with L$1_2$ structure in combination with CoFe was shown to provide $J_\mathrm{eff} > 1\,\mathrm{erg}/\mathrm{cm}^2$. However, ordered Mn$_3$Ir requires careful temperature control during deposition and needs an additional high temperature annealing.\cite{Tsunoda2006}

Within the effective Meiklejohn-Bean model\cite{RaduZabel} one can derive the expression $K_\mathrm{eff} = J_\mathrm{eff} / t_\mathrm{MnN}^\mathrm{crit}$ for the effective uniaxial anisotropy constant $K_\mathrm{eff}$, with the critical MnN thickness  $t_\mathrm{MnN}^\mathrm{crit}$ below which no exchange bias is observed. Evaluating this expression with $ t_\mathrm{MnN}^\mathrm{crit} \approx 11\,\mathrm{nm}$ gives  $K_\mathrm{eff} =  3.7\cdot 10^5\,\mathrm{erg}/\mathrm{cm}^3 = 37\,\mathrm{kJ}\,/\,\mathrm{m}^3$ or $K_\mathrm{eff} \approx 4.4\,\mu\mathrm{eV}$ per Mn atom. This anisotropy value is nearly three times larger than the anisotropy of FeMn ($K_\mathrm{eff} =  1.35 \cdot 10^5\,\mathrm{erg}/\mathrm{cm}^3$),\cite{Mauri87} but is slightly smaller than the anisotropy of NiMn ($K_\mathrm{eff} =  5 \cdot 10^5\,\mathrm{erg}/\mathrm{cm}^3$) and much smaller than the anisotropy of IrMn ($K_\mathrm{eff} =  2 \cdot 10^6\,\mathrm{erg}/\mathrm{cm}^3$) at room temperature.\cite{Carey01} From the data cited in Ref. \onlinecite{Rickart2005} for PtMn one can extract  $K_\mathrm{eff} = J_\mathrm{eff} / t_\mathrm{MnN}^\mathrm{crit} \approx 6.7 \cdot 10^5\,\mathrm{erg}/\mathrm{cm}^3$, which is again somewhat larger than the anisotropy of MnN.

\subsubsection{Influence of the Deposition pressure, Annealing Temperature, and the Nitrogen Content}

\begin{figure}[t]
\includegraphics[width=8.6cm]{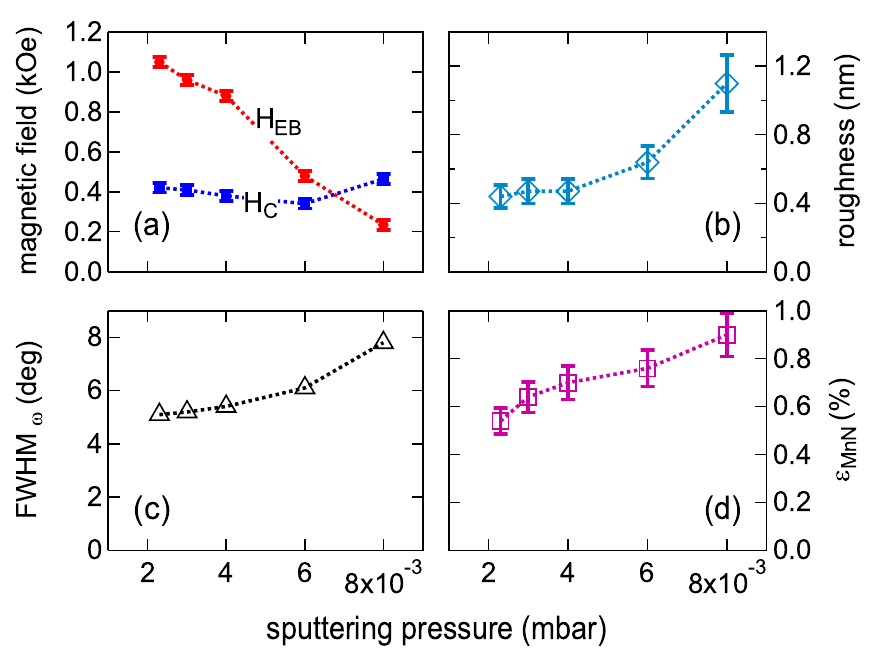}
\caption{Deposition pressure dependence of (a): exchange bias and coercive field ($t_\mathrm{MnN} = 30\,\mathrm{nm}$ $t_\mathrm{CoFe} = 1.8\,\mathrm{nm}$, annealing at 325$^\circ$C, 15\,min). (b): Surface roughness (root mean square) from x-ray reflectivity, (c): rocking curve FWHM of the (002) x-ray reflection, and (d): microstrain parameter $\varepsilon_\mathrm{MnN}$ for $t_\mathrm{MnN} = 30\,\mathrm{nm}$ without annealing.}
\label{fig:pressure_moke_xrd}
\end{figure}

In Figure \ref{fig:pressure_moke_xrd}\,(a) we show the dependence of the exchange bias and coercive field on the total pressure during deposition with the partial pressure ratio kept fixed at $p_\mathrm{N2} / (p_\mathrm{N2} + p_\mathrm{Ar} ) = 50\%$. The coercive field does not significantly change with the deposition pressure, however the exchange bias shows a strong, nearly linear decrease with the deposition pressure between $2 \cdot 10^{-3}$ and $8 \cdot 10^{-3}$\,mbar. This can be traced back to a deterioration of the MnN film growth as shown in Figure \ref{fig:pressure_moke_xrd}\,(b)-(d). The film roughness increases strongly, and also the rocking curve width and the microstrain increase to some extent at high deposition pressure. Remarkably, the lattice constant decreases by only $0.25\%$ (not shown), so the nitrogen content of the films does not seem to depend on the deposition pressure.

\begin{figure}[t]
\includegraphics[width=8.6cm]{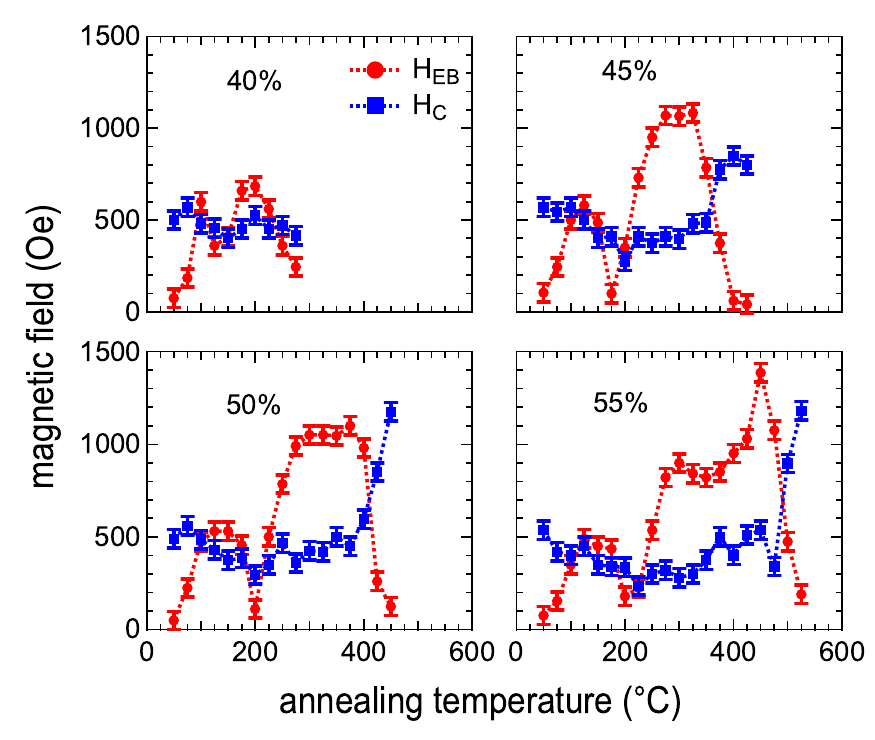}
\caption{Annealing temperature dependences of $H_\mathrm{EB}$ and $H_\mathrm{C}$ for different nitrogen partial pressure ratios during the reactive sputtering of MnN. The total pressure was held at $2.3 \cdot 10^{-3}$\,mbar. Film thicknesses were $t_\mathrm{CoFe} = 1.8$\,nm and $t_\mathrm{MnN} = 32$\,nm.}
\label{fig:heb_pressures}
\end{figure}

The exchange bias and coercive fields show a quite complex dependence on both the annealing temperature as well as on the nitrogen partial pressure during sputtering, see Figure \ref{fig:heb_pressures}. A high nitrogen partial pressure ratio between 45\% and 55\% is required to obtain high exchange bias. In all cases, maximum exchange bias is observed for annealing temperatures around $300^\circ$C. After annealing at much higher temperatures the exchange bias is lost and the coercive field enlarged at the same time. This may be related to nitrogen diffusion and destruction of the stack or to a structural or magnetic transition due to the loss of nitrogen. In favor of the second hypothesis speaks the fact that the annealing temperature at which exchange bias is lost increases with increasing nitrogen content in the films.

A particularly interesting feature of the annealing temperature dependence is the dip around $200^\circ$C, where the exchange bias almost vanishes. It is present for all nitrogen partial pressures. We assume that this dip is related to an irreversible structural or magnetic transition in the MnN or at the MnN\,/\,CoFe interface. Because of this complex behavior that is also related to crystal structure changes as discussed earlier, it is impossible to determine the blocking temperature of the stack from these annealing experiments. We investigate the blocking temperature distribution in the next section in detail.

Another remarkable feature of the annealing temperature dependence is the peak of the exchange bias after annealing of the 55\%-N$_2$ sample at $450^\circ$C. The increase of the exchange bias already sets in at $400^\circ$C, which is just above the N\'{e}el temperature of the MnN. As this is associated with a structural transition into the cubic phase, we suggest that after cooling through the transition in the external field, the MnN will recrystallize in a state that has enhanced coupling to the CoFe film and thereby generate a higher exchange bias. Additional experimental work will be necessary to clarify this.


\subsubsection{Blocking Temperature}
To gain a deeper understanding of the film thickness dependence of the exchange bias, we performed reversed field cooling experiments \cite{Nozieres2000} with different MnN thicknesses to obtain the blocking temperature distributions as a function of the MnN film thickness. Samples with $t_\mathrm{MnN} = 15,\,21,\,32,\,48$\,nm and $t_\mathrm{CoFe} = 1.8$\,nm were initially field cooled at 6.5\,kOe from $325^\circ$C, their hysteresis loops were measured, and then successively field cooled in a reversed field of 6.5\,kOe from $T_\mathrm{rev} = 50,\,75,\,\dots,\,300,\,325^\circ$C with the hysteresis loops measured between successive field cooling steps.

\begin{figure}[t]
\includegraphics[width=8.6cm]{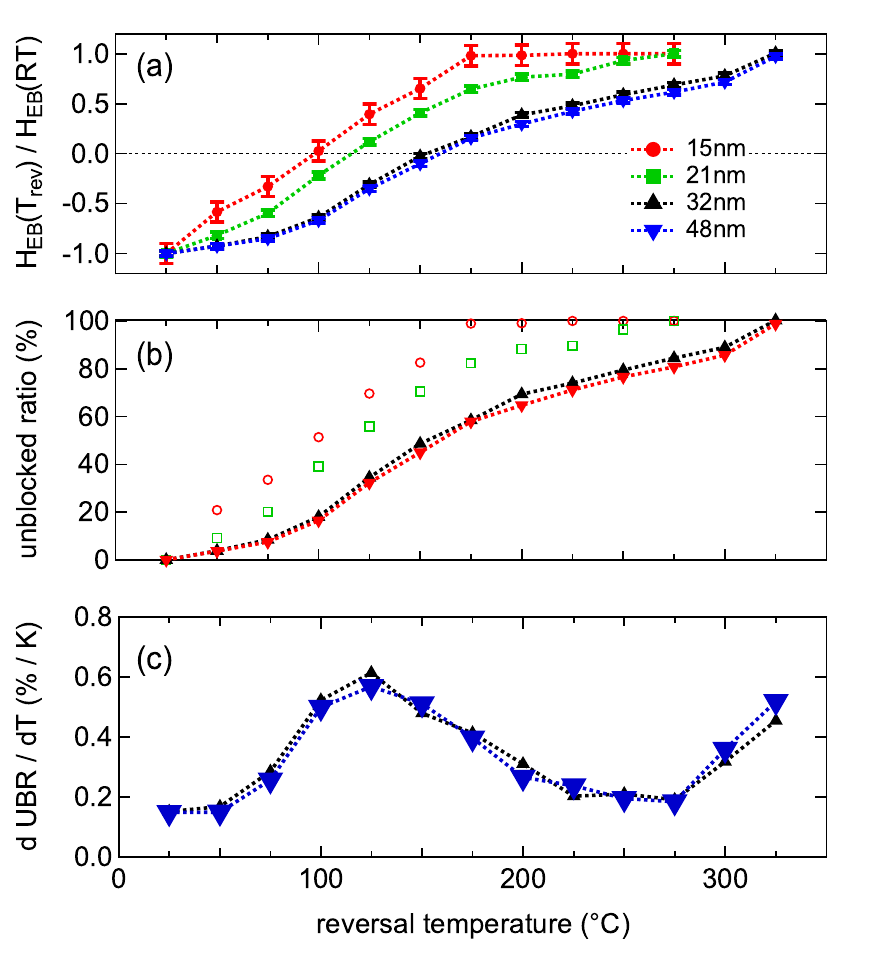}
\caption{Results from reversed field cooling experiments with four different MnN thicknesses. (a): Normalized exchange bias field $H_\mathrm{EB}$, (b): unblocked ratio, and (c): derivative of the unblocked ratio versus reversal temperature. Initial and reversal fields were 6.5\,kOe.}
\label{fig:blocking_temperatures}
\end{figure}

In Figure \ref{fig:blocking_temperatures} we present the results of this procedure. For easier comparison, we show the exchange bias normalized to the room temperature value in Figure \ref{fig:blocking_temperatures}\,(a). It is obvious that for thin MnN the blocking temperatures are much smaller than for larger MnN thickness. At $t_\mathrm{MnN} = 32$\,nm the blocking temperature distribution is essentially converged. The zero of this curve marks the median blocking temperature $\left< T_\mathrm{B} \right>$ of those antiferromagnetic grains that are still blocked at room temperature.  Clearly, this value varies strongly with the MnN film thickness and reaches a maximum of $\left< T_\mathrm{B} \right> = 160^\circ$C for large MnN thickness. The median blocking temperature is slightly higher compared to FeMn ($144^\circ$C), but lower compared to IrMn ($222^\circ$C), PtMn ($283^\circ$C), or NiMn ($355^\circ$C).\cite{Nozieres2000}

From the $H_\mathrm{EB}(T_\mathrm{rev})$ dependence we obtain the area fraction of unblocked grains, the so-called unblocked ratio $\mathrm{UBR}(T_\mathrm{rev})$ as
\begin{equation}
\mathrm{UBR}(T_\mathrm{rev}) = 100\% \cdot \left[ \frac{H_\mathrm{EB}(\mathrm{RT}) - H_\mathrm{EB}(T_\mathrm{rev})}{2 H_\mathrm{EB}(\mathrm{RT})} \right],
\end{equation}
which represents the cumulative distribution function of the blocking temperature (see Figure \ref{fig:blocking_temperatures}\,(b)). This quantity is only meaningful for the two films with large MnN thickness, because it refers to the ratio of grains that are unblocked at $T_\mathrm{rev}$ with respect to the number of grains that are still blocked at room temperature. The distributions however suggest that this number is already small for $t_\mathrm{MnN} = 15,\,21$\,nm and that the average blocking temperature is already below room temperature in these cases.

The blocking temperature distribution is obtained by taking the derivative $\partial\,\mathrm{UBR}(T)\,/\,\partial T$, which is shown in Figure \ref{fig:blocking_temperatures}\,(c). The blocking temperature distribution shows two maxima, one at around 125$^\circ$C and another one around the initial set temperature of 325$^\circ$C or more. This bimodal distribution is rather unusual for exchange bias systems\cite{Nozieres2000} and its origin is the subject of further studies we conduct. It may indicate the existence of at least two different populations of AFM/FM coupled grains with two different grain size distributions or different antiferromagnetic configurations.

\subsection{Applications}

\begin{figure}[t]
\includegraphics[width=8.6cm]{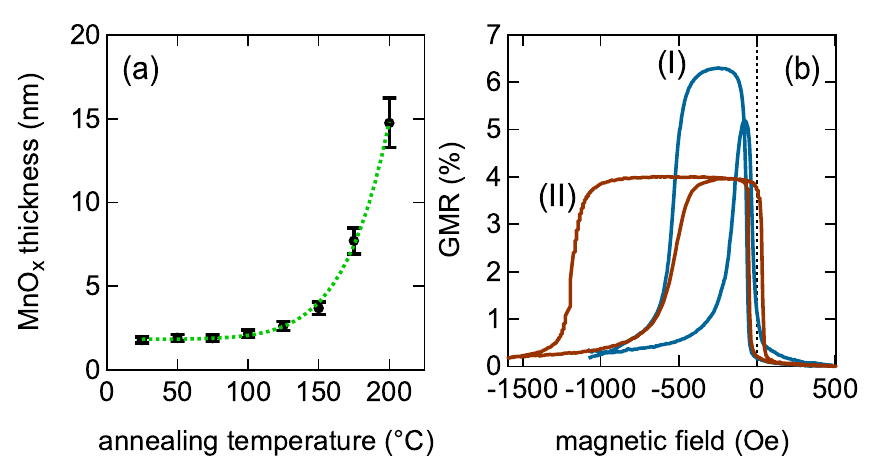}
\caption{(a): MnO$_x$ thickness after annealing a Ta 10 / MnN 30 bilayer in air for 1\,h. The fit represents Equation \ref{eq:diffusion}. (b): GMR loops at room temperature of (I): Ta 10 / MnN 30 / CoFe 2.2 / Cu 1.8 / CoFe 2.2 / Ta 0.5 / Ta$_2$O$_5$ 1.0 (250$^\circ$C, 15\,min) and (II): Ta 10 / MnN 30 / CoFe 2.2 / Cu 2.8 / CoFe 2.2 / Ta 0.5 / Ta$_2$O$_5$ 1.0 (325$^\circ$C, 15\,min).}
\label{fig:oxidation_and_gmr}
\end{figure}

If MnN is to be used for practical applications, such as for exchange biasing of TMR stacks, it has to withstand the lithography process; in particular, it has to be sufficiently stable against oxidation as to not degrade during air exposure after ion beam etching. To investigate the oxidation stability of the compound, we have made an oxidation study with x-ray reflectometry on a Ta / MnN film without any capping layer. The XRR shortly after deposition showed an oxide layer of 1.8\,nm, which did not change significantly after exposing the sample to air for 12\,h. Then samples were heated in air (relative humidity $\mathrm{RH} = 42\%$ at $24^\circ$C) up to $200^\circ$C for 1\,h. The results of this annealing experiment are shown in Figure \ref{fig:oxidation_and_gmr}\,(a).

Clearly, temperatures above 100$^\circ$C are necessary to promote the oxidation of the film. We model the oxygen migration into the film as a one-dimensional diffusion process. Here, we take the measured oxide thickness as
\begin{align}\label{eq:diffusion}
t_\mathrm{MnO}= \sqrt{\left< x^2(\tau) \right>} + t_0= \sqrt{2D\tau} + t_0\nonumber\\
\text{where}\quad D = D_0 \, \mathrm{e}^{-E_\mathrm{D}/k_\mathrm{B}T}.
\end{align}
Fitting the data with $\tau = 3600\,\mathrm{s}$ gives $t_0 = 1.8$\,nm, $D_0 = 2.2 \cdot 10^{-7}\,\mathrm{m}^2/\mathrm{s}$, and $E_\mathrm{D} = 1.23\,\mathrm{eV}$. So, at room temperature the diffusion coefficient is merely $D = 1.57 \cdot 10^{-28}\,\mathrm{m}^2/\mathrm{s}$. This appears surprising at a first glance, since it is well known that Mn compounds oxidize easily in many cases. To understand the good oxidation stability of MnN we have to consider that MnN is a densely packed structure with the small N atoms residing in the (nearly) octahedral interstices of the fct Mn lattice, giving rise to a high density of the material. Evaluating the Pilling-Bedworth ratio $\mathrm{PBR} = V_\mathrm{MnO}/V_\mathrm{MnN} = 1.16$, we find that it is close to 1, which means that a stable and only weakly strained, self-passivating oxide layer is formed on the surface of MnN that protects the bulk material from oxidation.\cite{Pilling1923} In summary, we expect MnN to be well suitable for lithographic processing.

Finally, as a proof-of-principle for integration of MnN into spintronic devices, we made current-in-plane GMR spin-valve stacks with  Ta 10 / MnN 30 / CoFe $t_\mathrm{CoFe}$ / Cu $t_\mathrm{Cu}$ / CoFe $t_\mathrm{CoFe}$ / Ta 0.5 / Ta$_2$O$_5$ 1.0. It is well known that Cu diffuses strongly at temperatures above $300^\circ$C, which are required to establish a high exchange bias with MnN. Nevertheless, by varying $t_\mathrm{CoFe}$ and $t_\mathrm{Cu}$ we obtained GMR stacks that showed exchange bias up to 850\,Oe, clear separation of the parallel and antiparallel magnetic states or a GMR amplitude up to 6.3\%, see Figure \ref{fig:oxidation_and_gmr}\,(b). Thus, it is clear that MnN can be used as an exchange biasing material in spintronic devices.

\section{Summary}

We have prepared Ta / MnN / CoFe exchange bias stacks by reactive magnetron sputtering and optimized the deposition conditions carefully. From our structural analysis we infer that a novel phase, $\theta'$-MnN with $c/a \approx 1.04$, was grown with $(001)$-fiber texture. Our findings demonstrate that MnN is a useful antiferromagnet for exchange biasing in spinelectronic devices. The MnN/CoFe system has large exchange bias (up to 1800\,Oe) at room temperature, reasonably large blocking temperature ($\left< T_\mathrm{B} \right> = 160^\circ$C) and is cheap in terms of materials and processing cost.

\acknowledgements
We thank the Ministerium f\"ur Innovation, Wissenschaft und Forschung des Landes Nordrhein-Westfalen (MIWF NRW) for financial support. We further thank G\"unter Reiss for making available laboratory equipment.


\begin{thebibliography}{50}
\bibitem{Gregg2002} J. F. Gregg, I. Petej, E. Jouguelet, and C. Dennis, J. Phys. D: Appl. Phys. \textbf{35}, R121 (2002).
\bibitem{Chappert2007} C. Chappert, A. Fert, and N. Van Dau, Nature Materials \textbf{6}, 813 (2007).

\bibitem{Meiklejohn1957} W. H. Meiklejohn and C. P. Bean, Phys. Rev. Lett. \textbf{105}, 904 (1957).
\bibitem{Nogues1999} J. Nogu\'{e}s and I.K. Schuller, J. Magn. Magn. Mater. \textbf{192}, 203 (1999).
\bibitem{Berkowitz1999} A. E. Berkowitz and K. Takano, J. Magn. Magn. Mater. \textbf{200}, 552–570 (1999).
\bibitem{Stamps2000} R. L. Stamps, J. Phys. D. Appl. Phys. \textbf{34}, 444 (2001).
\bibitem{Kiwi2001} M. Kiwi, J. Magn. Magn. Mater. \textbf{234}, 584 (2001).
\bibitem{Ohldag2003} H. Ohldag, A. Scholl, F. Nolting, E. Arenholz, S. Maat, A. Young, M. Carey, and J. Stöhr, Phys. Rev. Lett. \textbf{91}, 017203 (2003).
\bibitem{OGrady2010} K. O\'{}Grady, L. E. Fernandez-Outon, and G. Vallejo-Fernandez, J. Magn. Magn. Mater. \textbf{322}, 883 (2010).

\bibitem{Nozieres2000} J. P. Nozi\`{e}res et al., J. Appl. Phys. \textbf{87}, 3920 (2000).
\bibitem{Ali2003} M. Ali et al., Phys. Rev. B \textbf{68}, 214420 (2003).
\bibitem{Rickart2005} M. Rickart, A. Guedes, J. Ventura, J. B. Sousa, and P. P. Freitas, J. Appl. Phys. \textbf{97}, 10K110 (2005).

\bibitem{Glaister2010} B. J. Glaister and G. M. Mudd, Minerals Engineering \textbf{23}, 438 (2010).

\bibitem{Gokcen1990} N. A. Gokcen, Bull. Alloy Phase Diagr. \textbf{11}, 33 (1990).

\bibitem{Otsuka1977} N. Otsuka, Y. Hanawa, and S. Nagakura, Phys. Status Solidi (a) \textbf{43}, K127 (1977).
\bibitem{Suzuki2000} K. Suzuki, T. Kaneko, H. Yoshida, Y. Obi, H. Fujimori, and H. Morita, J. Alloys Compd. \textbf{306}, 66 (2000).
\bibitem{Leineweber2000} A. Leineweber, R. Niewa, H. Jacobs, and W. Kockelmann, J. Mater. Chem. \textbf{10}, 2827 (2000).
\bibitem{Suzuki2001} K. Suzuki, Y. Yamaguchi, T. Kaneko, H. Yoshida, Y. Obi, H. Fujimori, and H. Morita, J. Phys. Soc. Jpn. \textbf{70}, 1084 (2001).

\bibitem{Lambrecht2003} W. Lambrecht, M. Prikhodko, and M. Miao, Phys. Rev. B \textbf{68}, 174411 (2003).

\bibitem{Yang02} H. Yang, H. Al-Brithen, E. Trifan, D. C. Ingram, and A. R. Smith, J. Appl. Phys. \textbf{91}, 1053 (2002).

\bibitem{Williamson1953} G. K. Williamson and W. H. Hall, Acta Metall. \textbf{1}, 22 (1953).

\bibitem{RaduZabel} F. Radu and H. Zabel, Magnetic Heterostructures, Exchange Bias Effect of Ferro-/Antiferromagnetic Heterostructures, Springer Tracts in Modern Physics Volume 227 (Springer Berlin Heidelberg, 2008).

\bibitem{Tsunoda2006} M. Tsunoda, K. Imakita, M. Naka, and M. Takahashi, J. Magn. Magn. Mater. \textbf{304}, 55 (2006).
\bibitem{Reck69} R. A. Reck and D. L. Fry, Phys. Rev. \textbf{184}, 492 (1969).

\bibitem{Mauri87} D. Mauri, E. Kay, D. Scholl, and J. K. Howard, J. Appl. Phys. \textbf{62}, 2929 (1987).
\bibitem{Carey01} M. J. Carey, N. Smith, B. A. Gurney, J. R. Childress, and T. Lin, J. Appl. Phys. \textbf{89}, 6579 (2001).

\bibitem{Pilling1923} N.B. Pilling, R. E. Bedworth, J. Inst. Met. \textbf{29}, 529 (1923).

\end{thebibliography}
\end{document}